\documentclass[a4paper,fleqn,usenatbib]{mnras}

\usepackage[T1]{fontenc}
\usepackage{ae,aecompl}
\usepackage{epsf,rotating}
\usepackage{amsmath}
\usepackage[]{hyperref}
\usepackage{amssymb}
\usepackage{color}
\usepackage{verbatimbox}
\usepackage{bm}
\usepackage{bigfoot}
\usepackage{cleveref}
\newcommand{\msolar}{\;{\rm M}_{\odot}}
\newcommand{\fesc}{f_{\mathrm{esc}}}
\newcommand{\fescs}{f_{\mathrm{esc,\star}}}
\newcommand{\fesca}{f_{\mathrm{esc,AGN}}}

\newcommand{\rion}{{R_{\rm ion}}}
\newcommand{\ragn}{{R_{\rm ion,AGN}}}
\newcommand{\rst}{{R_{\rm ion,\star}}}
\newcommand{\rrec}{{R_{\rm rec}}}

\title[AGN contribution to EoR]{Constraining the contribution of active galactic nuclei to reionisation}

\author[S. Hassan et al.]{
\parbox[t]{\textwidth}{\vspace{-1cm}
Sultan Hassan$^{1,2}$\thanks{E-mail:sultanier@gmail.com}, Romeel Dav\'e$^{1,3,4,5}$, Sourav Mitra$^{6}$, Kristian Finlator$^7$, Benedetta Ciardi${^2}$, Mario G. Santos$^{1,8}$}
\\
\\$^1$ University of the Western Cape, Bellville, Cape Town, 7535, South Africa
\\$^2$ Max-Planck-Institut f\"ur Astrophysik, Karl-Schwarzschild-Strasse 1, D-85748 Garching b. M\"unchen, Germany
\\$^3$ South African Astronomical Observatories, Observatory, Cape Town, 7925, South Africa
\\$^4$ African Institute for Mathematical Sciences, Muizenberg, Cape Town, 7945, South Africa
\\$^5$ Center for Computational Astrophysics, Simons Foundation, New York, New York
\\$^6$ Surendranath College, Dept. of Physics, 24/2 M. G. Road, Kolkata 700009, India
\\$^7$ New Mexico State University, Las Cruces, NM 88003, United States
\\$^8$ SKA SA, The Park, Park Road, Pinelands 7405, South Africa
}
\date{Accepted XXX. Received YYY; in original form ZZZ}

\pubyear{2017}

\begin{document}
\label{firstpage}
\pagerange{\pageref{firstpage}--\pageref{lastpage}}
\maketitle

 \begin{abstract}
Recent results have suggested that active galactic nuclei (AGN)
could provide enough photons to reionise the Universe.
We assess the viability of this scenario using a semi-numerical
framework for modeling reionisation, to which we add a quasar
contribution by constructing a Quasar Halo Occupation Distribution
(QHOD) based on Giallongo et al. observations.  Assuming
a constant QHOD, we find that an AGN-only model cannot simultaneously
match observations of the optical depth $\tau_e$, neutral fraction, and ionising
emissivity.  Such a model predicts $\tau_e$ too low by $\sim 2\sigma$
relative to Planck constraints, and reionises the Universe at $z\la
5$.  Arbitrarily increasing the AGN emissivity to match these results
yields a strong mismatch with the observed ionising emissivity at
$z\sim 5$.  If we instead assume a redshift-independent AGN luminosity
function yielding an emissivity evolution like that assumed in
Madau \& Haardt model, then we can match $\tau_e$ albeit with late
reionisation; however such evolution is inconsistent with observations
at $z\sim 4-6$ and poorly motivated physically.  These results arise
because AGN are more biased towards massive halos than typical
reionising galaxies, resulting in stronger clustering and later
formation times. AGN-dominated models produce larger ionising
bubbles that are reflected in $\sim\times 2$ more 21cm power on all
scales. A model with equal parts galaxies and AGN contribution is
still (barely) consistent with observations, but could be
distinguished using next-generation 21cm experiments HERA and
SKA-low.  We conclude that, even with recent claims of more faint
AGN than previously thought, AGN are highly unlikely to dominate
the ionising photon budget for reionisation.
\end{abstract}




\section{Introduction}

The nature of the sources driving the epoch of reionisation (EoR) in
the early Universe remains uncertain.  
It is canonically believed that star-forming galaxies have provided
the bulk of the ionising photon budget required to complete
reionisation~\citep{bar01,loeb01}.  This is because there is a
significant decrease of observed active galactic nuclei (AGN) candidates at redshifts $z>3$,
such that the contribution from star-forming galaxies is expected
to well exceed that of AGN at $z>6$~\citep{shgi87,shanmat07,hop07,glik11,mas12,hama12,miche17,ricci17}.  However,
there remain large uncertainties in the contribution of both
star-forming galaxies and AGN to reionisation.  Current constraints
are now consistent with a minimal contribution from very low
metallicity Population III stars~\citep[e.g.][]{rob15}, but
there is still the issue of the highly uncertain ionising photon
escape fraction $\fescs$.  Direct observations of $\fescs$ are quite
difficult at $z\ga 4$ owing to the ubiquity of strong absorption
systems that suppress Lyman continuum flux and the difficulty in
removing foreground interlopers, but careful measurements generally
indicate $\fescs$ less than a few
percent~\citep[e.g.][]{graz16,vas16}, with some evidence for
a higher $\fescs$ in lower-mass galaxies~\citep{van16,graz17,bian17}.

Theoretical models have tried to constrain $\fescs$ indirectly by
matching models to other data, making a variety of assumptions for
$\fescs$ such as a constant (e.g.  \citealt{rob13,fink15,ma15,hassan16}),
redshift-dependent (e.g.  \citealt{KuFa12,mit13,fin15,qin17}),
mass-dependent (e.g.  \citealt{gne07,Yaj11,wise14,paa15}), and
recently UV magnitude-dependent $\fescs$~\citep{anderson13,jape17} in order
to match simultaneously various reionisation constraints.  The
currently-favoured lower value of Thomson scattering integrated
optical depth ($\tau = 0.058\pm 0.012$) measured by~\citet{planck16}
prefers rather sudden and late reionisation scenarios, which relaxes
the previously stringent constraints on the ionising photon budget.
In~\citet{hassan17}, we performed a detailed Monte Carlo Markov
Chain (MCMC) analysis to constrain our semi-numerical model to
several EoR key observables and found that $\fescs$ is highly degenerate
with the ionising emissivity amplitude, leading to a best fit value
of $\fescs=0.25^{+0.26}_{-0.13}$, which allows a substantial range
but is generally higher than available (lower-redshift) observations.
Without a firmer understanding or direct measurement of $\fescs$,
it is difficult to conclusively argue that star-forming galaxies
can provide all the photons required for reionisation.

Recently, there has been renewed interest in assessing the contribution
of AGN to the reionising photon budget.  Previous estimates of the
AGN contribution relied on an extrapolation to faint luminosities
based on lower-redshift results.  But recent deep observations have
enabled a more direct characterisation of the faint end.  \citet[hereafter G15]{gia15}
identified 22 faint AGN candidates at $z>4$ and inferred a significantly
steeper faint-end slope than what is seen at lower redshifts.  We
note that claims of such a steep faint end remain controversial;
for instance \citet{parsa17} was unable to confirm a substantial
fraction of the G15 candidates based on additional
multi-wavelength data. Furthermore, recent spectroscopic
surveys~\citep{kim15,jia16} have concluded that the observed quasars population at high redshift
might not be enough to fully reionise the Universe. Nonetheless, the
differing claims 
have led to speculation that AGN could provide the primary ionising 
photon contribution in order to keep the inter-galactic medium (IGM)
highly ionised~\citep[e.g.][]{maha15}.  These claims further favor a late reionisation scenario in which
the flatness observed in the ionising emissivity measurements by~\citet{bec13} might arise naturally. In addition, they might also support
the early and extended Helium reionisation observed by~\citet{wor16}. Independently, \citet{char17} argued 
that the large scale opacity fluctuations in the Ly$\alpha$ forest 
measured by~\citet{bec15} could be explained if AGN dominate the 
ionising UV background at $z\sim6$ (see also~\citealt{char15}). Hence the contribution of AGN 
to reionisation remains uncertain and potentially important or even dominant.

The idea that AGN might have driven cosmic reionisation has so far
been investigated mostly in terms of global quantities, such as the
ability to match the optical depth or comoving ionising emissivity
constraints \citep[e.g.][]{maha15,mit16,makim16,khair16,qin17}.  It remains to be
demonstrated whether AGN-driven models are able to simultaneously
satisfy all the current reionisation-epoch constraints.  An important
upcoming addition to the pantheon of constraints will be the 21cm
EoR power spectrum, which may be substantially different for AGN-
versus star formation-driven reionisation, if AGN and star-forming
galaxies cluster in different ways as one might naively expect.  An
early attempt by~\citet{gewy09} to assess the effect of AGN on the
21cm power spectrum using a seminumerical scheme concluded that the
effect is likely to be small, but more recent semi-numerical models
by \citet{kul17} have suggested the opposite, that AGN
produce signficantly different 21cm signal.  However, \citet{kul17} populates AGN
only in the most massive halos using abundance matching to the halo
velocity, employing the observed velocity-black hole mass relation
at lower redshifts~\citep{tre02,fer02}, which thus effectively
adopts a unity duty cycle of AGN for massive halos. However, recent
results from Hyper Suprime-Cam suggest that quasars do not necessarily
live in the most overdense regions where massive halos are expected
to reside, and that their duty cycle is below a few
percent~\citep{he17}. Accounting for sub-unity duty cycles inevitably drives 
black holes into lower-mass halos, altering the implied emissivity associated
with haloes and epochs where they are not directly-measured. Moreover, the G15 data
suggest that the AGN driving reionisation are rather faint, which
may not be associated with the most massive halos.  Without
a proper treatment for AGN occupancy (duty cycle) and a more comprehensive
analysis of all the implications of AGN-driven reionisation, it is
difficult to properly assess the viability of this scenario.

In this paper, we build on our semi-numerical framework based on
the {\sc SimFast21} code to evolve the EoR ionisation field, which
allows us to examine a range of EoR observations as we have done
in \citet{hassan16,hassan17}.  To explore the AGN contribution, we
populate AGN into halos with  a more physically motivated approach that
utilises both the observed luminosity function and abundance matching,
thereby generating a Quasar Halo Occupancy Distribution (QHOD); our
scheme partially follows the recipe summarized in~\citet{chofer05}.
We constrain our QHOD to match the G15 AGN LF fit at
$z=5.75$, and assign AGN randomly into halos.  This QHOD {\it
predicts} a duty cycle that is close to unity for extremely massive
halos, but drops to sub-percent values at intermediate halo masses.
To obtain the AGN emissivity, we utilise the strong correlation
observed between the circular velocity and black hole mass following
low redshift observations~\citep{tre02,fer02}.  We account for this
additional AGN photon contribution while we evolve our {\sc SimFast21}
density and ionisation field, including the effects of recombination
and time-evolving neutral fractions.

This work improves on previous efforts in several ways. First, using
our {\sc SimFast21}-based framework, we examine a wider variety of
simultaneous constraints on the evolution of AGN-driven reionisation,
including the Thomson optical depth, the mean cosmic neutral fraction
evolution, and the ionising emissivity at the end of reionisation.
Second, our model for populating quasars into halos is more realistic
than previous works because we apply constraints beyond just abundance
matching, allowing us to directly constrain the duty cycle of AGN
as a function of halo mass.  Third, we forecast upcoming 21cm EoR
power spectrum measurements from LOFAR, HERA, and SKA, and illustrate
how such future data might be able to constrain the fractional
contribution of AGN to reionisation.  Our primary conclusion is
that it is very difficult to reconcile purely AGN-driven reionisation
based on the (optimistic) G15 AGN luminosity function measurements
with current global reionisation constraints.  Future 21cm data
should provide a new avenue to more precisely characterise the
contribution of AGN to reionisation.

This paper is organized as follows: In Section~\ref{sec:sims}, we
describe our semi-numerical simulation and the AGN model
implementation and calibration.
We compare AGN with star-forming galaxies models in terms of 
their EoR observables, present the 21cm predictions, and discuss how future experiments 
can discriminate between these models in Section~\ref{sec:impacts}.
We finally conclude in Section~\ref{sec:con}. Throughout this work, we adopt a $\Lambda$CDM cosmology in which
$\Omega_{\rm M}=0.3$, $\Omega_{\rm \Lambda}=0.7$, $h\equiv H_0/(100
\, \mathrm{km/s/Mpc})=0.7$, a primordial power spectrum index
$n=0.96$, an amplitude of the mass fluctuations scaled to $\sigma_8=0.8$,
and $\Omega_b=0.045$. We quote all results in comoving units, unless
otherwise stated.

\section{Simulations Using {\sc SimFast21}}\label{sec:sims}

We use the recently developed Time-integrated version of our
semi-numerical code {\sc SimFast21}~(\citealt{san10}) that has been
presented in ~\citet{hassan17}. We here briefly review the simulation
and defer to ~\citet{san10} for full details about the basic
algorithm, and to ~\citet{hassan16,hassan17} for more information
about our subsequent improvements.

The dark matter density field is generated using a Monte-Carlo
Gaussian approach, which is dynamically evolved into the non-linear
regime via applying the~\citet{zeldovich70} approximation. The dark
matter halos are generated using the well known excursion set
formalism (ESF). In the Time-integrated model, the ionised regions are
identified using a similar form of the ESF that
is based on comparing the time-integrated ionisation rate $\rion$
with that of the recombination rate $\rrec$ and the local neutral Hydrogen
density within each spherical volume specified by the ESF.  Regions are flagged as ionised if:
\begin{equation}
\label{eq:new_con}
\int\fesc \rion \,\, dt \geq \int x_{\rm HII}\, \rrec\,\, dt +  (1 - x_{\rm HII})\, N_{\rm H}\, ,
\end{equation}
where $\fesc$ is the photon escape fraction, $x_{\rm HII}$ is the
ionised fraction, and $N_{\rm H}$ is the total number of hydrogen
atoms. This is a generalized form of the ionisation condition in
the Time-integrated model, which can be used for any ionising source
or sink populations to run the reionisation calculations.  With
this ionisation condition, reionisation occurs more suddenly compared
to our previous Instantaneous model developed in~\citet{hassan16},
in which the ionisation condition was based on an instantaneous
comparison of $\rion$ and $\rrec$.  The more sudden reionisation is
favoured by recent~\citet{planck16} data, and in \citet{hassan17}
we showed that the Time-integrated ionisation condition produces
larger ionised bubbles, resulting in 21 power spectrum enhancement
on large scales.

\subsection{Sink model}

Reionisation, in short, is an evolving battle between ionising
photon sources and sinks.  To model sinks, we must account for the
clumping effects from small scales below what we can directly evolve
using the large-scale {\sc SimFast21} code (typically, sub-Mpc
scales).  We thus parametrize the inhomogeneous recombination rate
$\rrec$ from high-resolution full radiative transfer hydrodynamic
simulations (hereafter 6/256-RT)~\citep{fin15} as a function of
over-density $\Delta$ and redshift $z$, as follows:
\begin{equation}\label{eq:rrec}
\frac{\rrec}{V} =  A_{\rm rec}(1+ z)^{D_{\rm rec}}  \left[\frac{\left( \Delta/B_{\rm rec} \right)^{C_{\rm rec}}}{1+ \left( \Delta/B_{\rm rec} \right)^{C_{\rm rec}} } \right]^{4} \, , 
\end{equation}
where $A_{\rm rec} = 9.85 \times 10^{-24} $cm$^{-3}$s$^{-1}$ (proper
units), $B_{\rm rec} = 1.76 $, $C_{\rm rec}= 0.82$, $D_{\rm rec}=5.07$.
Consistent with ~\cite{sob14}, our recombination rate $\rrec$
parametrization suppresses the ionisation and 21cm power spectrum
on large scales.  Full details about the inhomogeneous
recombinations $\rrec$ parametrizations and impact on the EoR
observables can be found in~\citet{hassan16}. 

We note that AGN-only reionisation scenarios are found
to substantially heat the IGM~\citep{dalo16,ono17}, which lowers
the recombination rate. This may reduce $\rrec$ by up to a factor
of $\sim$ 2 in our AGN-only models, which in turn may slightly
advance reionisation by AGN, and hence improving the viability of
AGN-only models. We do not account for this effect in our calculation
since we expect it to be sub-dominant compared to other effects
related to halo growth, and here simply use the same sink model to
compare reionisation histories produced by Galaxies versus AGN.

\subsection{Source model: Star-forming galaxies}

For the stellar contribution, we use a parametrization obtained
from combining the 6/256-RT with larger-volume hydrodynamic galaxy
formation simulation~\citep{dav13} (hereafter 32/512), that have
both been shown to match a range of observations including
lower redshift data. From these simulations, we parametrize the
non-linear ionisation $\rst$ rate as a function of halo mass $M_{\rm
h}$ and redshift $z$ as follows:
\begin{equation}\label{eq:nion}
\frac{\rst}{M_{\rm h}} =  A_{\rm ion}\, (1 + z)^{D_{\rm ion}} \, ( M_{\rm h}/B_{\rm ion} )^{C_{\rm ion}} \, \exp\left( -( B_{\rm ion}/M_{\rm h})^{3.0} \right),
\end{equation}
where $A_{\rm ion} =1.08\times 10^{40}  \msolar^{-1} $s$^{-1}$,
$B_{\rm ion} = 9.51\times 10^{7}\msolar$, $C_{\rm ion} = 0.41$ and
$D_{\rm ion} = 2.28$. This ionisation rate is computed directly
from the star formation rate (SFR) of these simulations based
on stellar population models applied to star formation histories
of simulated galaxies.  

In \citet{hassan17} we considered a more generalized form of this
source model, and found that constraining these parameters against
several EoR observations using MCMC analysis resulted in best-fit
values that matched the above parameters to within uncertainties,
thereby validating the extrapolation from the small scales of
6/256-RT and 32/512 simulations to large scales covered by {\sc
SimFast21} simulations. We further showed that using this non-linear
ionisation rate relation boosts the small scales 21cm power spectrum
as compared with models assuming a linear relation between the
ionisation rate and halo mass; see \citet{hassan16, hassan17} for
more details.

\subsection{Source model: AGN}

The new aspect of the source model for this work is the AGN ionising
photon output.  We compute the ionisation rate from AGN $\ragn$
following partially the recipe summarized in~\citet{chofer05}.
Motivated by low redshift observations~\citep{tre02,fer02}, the
basic assumption is that the black hole mass $M_{\rm bh}$ is strongly
correlated with the hosting halo's circular velocity $ v_{\rm cir}$.
We assume that this correlation is independent of redshift and valid
during the reionisation redshifts. This correlation can be written
as:
\begin{equation}\label{eq:mbh}
\frac{M_{\rm bh}}{M_{\odot}} = A \, \left(\frac{v_{\rm cir}}{159.4\,{\rm km\,s^{-1}}} \right)^{5}\, , 
\end{equation}
where $A$ may be regarded as the black hole formation efficiency,
which is our only free parameter in the AGN source model at fixed
$\fesca$. We then fix $A$ to match the AGN ionising emissivity constraints
from G15, as we describe in \S\ref{sec:calib}. It is worthwhile
to mention that this observed correlation (Equation~\ref{eq:mbh}) would naturally arise if one applies
a self-regulation condition on the black hole growth as previously shown by~\citet{wylo03}.

For our adopted cosmology, the circular velocity $v_{\rm cir}$ of
a given halo mass $M_{h}$ is given by:
\begin{equation}\label{eq:vcir}
\frac{v_{\rm cir}}{{\rm km\,s^{-1}}} = 0.014\left(   M_{h} \sqrt{\Omega_{m}(1+z)^{3} + \Omega_{\Lambda}}\right)^{1/3}.
\end{equation}
Having obtained the black hole mass $\rm M_{\rm bh}$, the Eddington
luminosity in the B-band is given by~\citep{chofer05}
\begin{equation}\label{eq:lb}
\frac{L_{B}}{L_{\odot,B}} = 5.7\times 10^{3}\, \frac{M_{\rm bh}}{M_{\odot}}\,.
\end{equation}

Given this B-band luminosity, we must now determine the ionising photon output.
Following~\citet{scbu03} and~\citet{tel02}, we assume the spectral energy distribution for 
AGN takes a power law form:
\begin{equation}\label{eq:lnu}
L_{\nu} = L_{912} \left(\frac{\nu}{\nu_{912}}\right)^{-1.57},
\end{equation} 
where $\rm L_{912}$ is the luminosity at the Lyman limit that is given by:
\begin{equation}\label{eq:l912}
\frac{L_{912}}{\rm ergs\, s^{-1}\, Hz^{-1}} = 10^{18.05}\, \frac{L_{B}}{L_{\odot,B}}\,.
\end{equation}
We then integrate the above over all frequencies to find the ionisation rate $\ragn$ as follows:
\begin{equation}\label{eq:ragn}
\ragn = \int_{\nu_{912}}^{\infty} \, \frac{L_{\nu}}{h\nu} \, d\nu =  \frac{L_{912}}{1.57\, h}.
\end{equation}
This explains how we compute the AGN ionisation rate $\ragn$ given the host halo properties.

Next, we must populate the AGN into our halos.  Here is where we
make use of the G15
AGN Luminosity Function (LF).  G15 evaluated this at $\lambda =
1450 {\rm \r{A}}$ for several redshifts higher than $z = 4$. We
then use~G15 LF fit at their highest redshift z=5.75 to
compute the number of AGN as a function of halo mass in our
simulations. G15 LF at z=5.75 can be best fit using a
double power law as follows:
\begin{equation}\label{eq:lf}
\phi = \frac{\phi^{*}}{10^{0.4(M_{\rm break}-M)(\beta - 1.0)} + 10^{0.4(M_{\rm break}-M)(\gamma - 1.0)} }\, , 
\end{equation}
where $\phi$ is the comoving AGN density, $M$ is the absolute magnitude computed via the standard relation $M_{\rm AB} = -2.5\, {\rm log}_{10} (L_{\nu}) + 51.60$, log$_{10}$ $\phi^{*}=-5.8$ Mpc$^{-3}$, $M_{\rm break}=-23.4$, $\beta=1.66$, and $\gamma=3.35$.

Putting this together, our procedure to populate AGN into halos is as follows:
\begin{enumerate}
\item We bin the halo catalogs as a function of halo mass, and find 
the average halo mass in every bin of $\Delta\log_{10} M_{\rm h} = 0.34$.
\item Using~\Crefrange{eq:mbh}{eq:lnu}, we compute the corresponding AGN 
$L_{1450}$ and $M_{1450}$ of each halo mass bin.
\item We then obtain the number of AGN for each halo mass bin using Equation~\eqref{eq:lf}, 
which turns out to always be less than the actual number of halos; the ratio of these
numbers is the duty cycle of AGN for that halo mass bin.
\item We randomly assign the appropriate number of AGN into halos within that mass bin.
\end{enumerate}

Note that in step (iii) one ideally may assume Poisson fluctuations around the number of AGN following~\citet{mcq09} QSO Method I, since the luminosity function, in principle, yields the average number of AGN at a given magnitude bin per the simulation volume. We ignore these fluctuations for two complementary reasons. First, the number of AGN obtained from the G15 LF is very large, particularly, at the faint end (N $\sim$ 10$^{6}$) around which the Poisson fluctuations can be neglected ($ \sqrt{N}  \sim$ 10$^{3}$). Second, as will be seen later, our results are mainly driven by the strong AGN clustering at the faint end, and hence adding Poisson fluctuations at the very bright end (e.g. first few magnitude bins) is unlikely to affect the results since bright sources are rare. In such a situation,  the average number of AGN is a very good approximation to the actual number of AGN. We then round off the resulting AGN number in order not to populate halos with fractional AGN, but this in fact is a small correction.

Following the above procedure, we now have plausible AGN population in our simulation box at
$z=5.75$. To quantify the evolution of the AGN population, we must
make a choice regarding the evolution of AGN relative to that of
the halos.  Given that theoretical predictions for AGN evolution
are relatively uncertain, the simplest assumption is to assume that
the relationship between AGN and their host halos do not change at
$z\geq 5.75$.  In other words, we assume that the Quasar Halo
Occupation Distribution (QHOD) is non-evolving.  This is a reasonable
assumption since the HOD of galaxies have been studied extensively
(see \citealt{yosh01,ber03}, and references therein) and it is found
that the HOD is nearly a redshift-independent quantity.

\begin{figure}
\centering
\setlength{\epsfxsize}{0.5\textwidth}
\centerline{\includegraphics[scale=0.5]{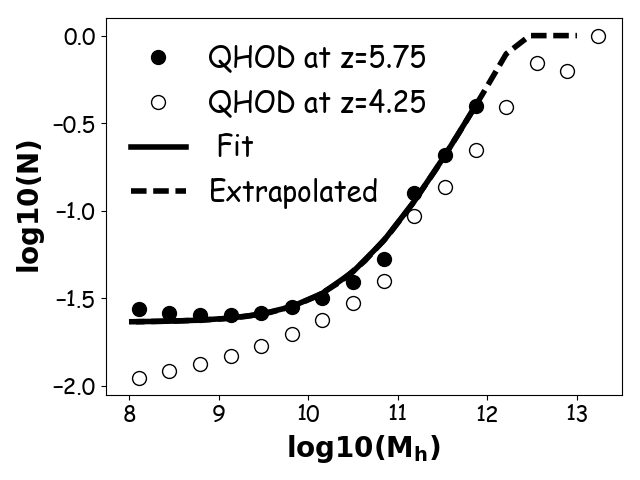}}
\caption{The Quasar Halo Occupancy Distribution (QHOD) as function of halo mass $M_{h}$ computed from~G15 LF at z=5.75 (Equation~\ref{eq:lf}, closed circles) and at z=4.25 (open circles). The QHOD is relatively similar at these two redshift bins, providing an evidence that the QHOD doesn't evolve strongly with redshift. The solid line represents the fitting function written in Equation~\eqref{eq:hod} for QHOD data at z=5.75. The fitting function is extrapolated (dashed) for halo masses higher than $M_{h} = 10^{12}$ and set to unity as an extreme occupation condition since AGN number should not exceed halos number. The QHOD increases as the $M_{h}$ increases, showing that there are few AGN at massive halo mass bins. This fitting function will be used to evolve AGN from z=5.75 to high redshifts in our constant QHOD AGN fiducial model. }
\label{fig:HOD}
\end{figure} 

The QHOD as a function of halo mass $M_{h}$ can directly be calculated
from G15 LF fit at z=5.75 (Equation~\ref{eq:lf}) as the ratio
between the number of AGN to that of their hosting halos for each
halo mass bin. Our QHOD can be well fit with a constant plus a power
law as follows:
\begin{equation} \label{eq:hod}
{\rm N } = \left(\frac{M_{h}}{2.19\times 10^{12}}\right)^{0.9} +  0.023\, .
\end{equation}
Note that the QHOD changes with different values of our free
parameter $A$ relating black hole mass to circular velocity, which
translates into a shift in magnitudes of the AGN.  Here we have
used $A=5\times 10^{5}$, a value at which our constant QHOD AGN
model is calibrated to reproduce the~G15 ionising emissivity
constraints, as will be discussed next in \S\ref{sec:calib}.

Figure~\ref{fig:HOD} shows the QHOD computed from~G15
observations at z=5.75 (closed circles) and our QHOD fit in Equation~\eqref{eq:hod}.
The QHOD represents a plausible description of the AGN occupancy
in their hosting halos.  Indeed, this can also be regarded as an
AGN duty cycle, if one (reasonably) postulates that every halo
contains a black hole but only some fraction of them are detectably
active.  The \citet{he17} observations suggest a duty cycle
of $0.001-0.06$ for moderate-mass halos, which is somewhat lower
than our model assumes but qualitatively agrees with the trend
that the duty cycle is smaller in lower-mass halos.  For comparison,
we also plot the QHOD at $z=4.25$ that is computed from~G15
LF at that redshift bin (open circles in Figure~\ref{fig:HOD}). We
notice that the QHOD data at $z=5.75$ and $z=4.25$ are fairly
similar, differing by $\sim 30\%$ for all $M_h\ga 10^{10}M_\odot$.
This suggests that the QHOD doesn't evolve strongly with redshift,
and motivates us to fiducially assume that the QHOD does not evolve.
We will call this the ``constant QHOD" case. In this case,
we replace Equation~\eqref{eq:lf} with Equation~\eqref{eq:hod} in step (ii)
to compute directly the number of AGN in each halo mass bin at
higher redshifts. Note that in step (iv) AGN assignment is completely 
random and redshift-independent. As a result, halos with AGN may or
may not have AGN at the next time step. This is realistic since the 
simulation time step (dz=0.125) is typically larger than the
AGN lifetime.

As a counterpoint to this case, we also consider a model where the
AGN luminosity function is constant with time.  Here, we calculate
the AGN number at all redshifts based on~G15 LF fit at
$z=5.75$ (Equation~\ref{eq:lf}).  We will call this the ``constant
LF" case. This is less realistic because the QHOD here increases
strongly with redshift, since there are many fewer halos at higher
redshifts but the number of AGN remains fixed following the assumed constant LF.  Also, observations of the
AGN LF at lower redshifts ($z\la 6$) exhibit significant evolution,
so it seems unlikely that this evolution should suddenly cease.
Nonetheless, the emissivity evolution in this case turns out to be
similar to the AGN comoving ionising emissivity model assumed by
\citet{maha15}, hence it represents an interesting contrasting case
that we will examine in \S\ref{sec:impacts}.

\subsection{AGN source model calibration}\label{sec:calib}

\begin{figure}
\centering
\setlength{\epsfxsize}{0.5\textwidth}
\centerline{\includegraphics[scale=0.5]{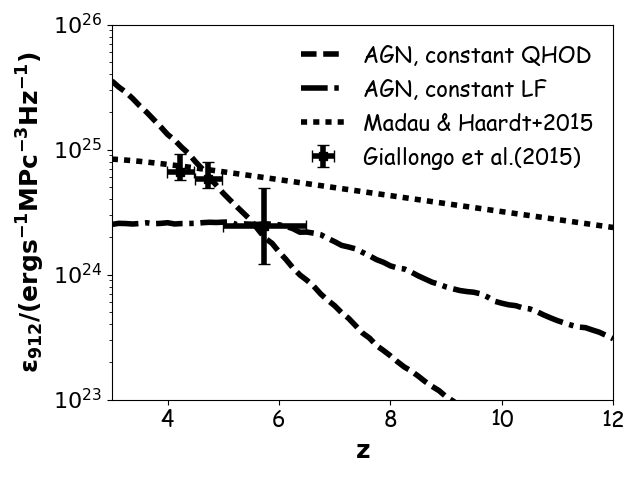}}
\caption{The comoving ionising emissivity of AGN at 912 \r{A}. The constant LF AGN model (dot-dashed) only matches the G15 constraints (1-$\sigma$ level) at z=5.75 which exhibits  a slowly growing emissivity evolution that is somewhat similar to the emissivity shape by~\citet{maha15} (dotted). The constant QHOD AGN model (dashed) yields more physical evolution as the ionising emissivity grows  rapidly which in turn results in matching almost all of G15 data.}
\label{fig:epsilon912}
\end{figure} 

Our next task is to calibrate the relationship between black hole mass and circular velocity via 
the normalization parameter A in Equation~\eqref{eq:mbh}. Observationally, this parameter is not 
tightly constrained and has only been measured at low redshifts.

We first calibrate the constant QHOD and  constant
LF AGN models to at least match the G15 ionising emissivity
constraints at z=5.75 in order to verify the possibility to complete
reionisation solely by AGN. This we achieve by tuning the black hole formation
efficiency $A$ in our AGN models to match the total ionising
emissivity measurements at 912 \r{A} ($\epsilon_{912} $),
which is the total escaped $L_{912}$ of all AGN divided by the
simulation comoving volume. The simulation configurations of these
models are presented in~\S\ref{sec:impacts} with the rest of our
fiducial models. We assume $\fesca = 100\%$ for AGN, which 
is standard~\citep[e.g.][]{maha15}.

We find that the  constant LF AGN model can match the~G15
ionising emissivity constraints with $A= 10^{6}$ whereas the
constant QHOD AGN model requires $A = 5\times10^{5}$. We
note that what is really constrained here is the product of $A\fesca$,
so we have the freedom to keep $A$ fixed and tune $\fesca$ instead;
all our results would be unchanged.  In this case, the  constant
QHOD and  constant LF AGN models would require $\fesca = 50,
100 \%$ at $A= 10^{6}$ to match the~G15 constraints,
respectively.  Note that we do make the assumption here that the
product $A\fesca$ does not vary with redshift. 

In Figure~\ref{fig:epsilon912}, we show the comoving ionising 
emissivity evolution obtained with the procedure discussed above. The constant LF AGN model produces a slowly
growing emissivity, which is similar to the evolution expected from
models in which ionising radiation is dominated by star-forming
galaxies, and similar in shape to that assumed in \citet{maha15}.  
While matching the~G15 constraints at $z=5.75$,
the  constant LF AGN model under-estimates the ionising
emissivity by a factor of $\sim$ 3 as compared with~G15
constraints at $z=4.75$ and $z=4.25$.  In contrast, the emissivity from our
fiducial constant QHOD AGN model matches
simultaneously~G15 data at several redshifts bins due to
the rapidly growing emissivity evolution as expected from an AGN
dominated model. This further validates our assumption that using 
constant QHOD is a more physically motivated approach than using
the constant LF.

Note that we have intentionally not applied a magnitude cut-off in
computing the integral of emissivity ($\epsilon_{912}$); G15
used a cutoff of $M_{1450} = -18$. At $z=5.75$, the total comoving
ionising emissivity is $\epsilon_{912} = 2.12\times 10^{24}$ erg
s$^{-1}$ Mpc$^{-3}$ Hz$^{-1}$, whereas with a magnitude cut-off of
$M_{1450} = -18$ it becomes $\epsilon_{912} = 1.58\times 10^{24}$ erg
s$^{-1}$ Mpc$^{-3}$ Hz$^{-1}$. This shows that those fainter AGN
contribute $\sim$ 25$\%$ to the total emissivity, which is modest
but not negligible.  We include this in order to check whether
including all faint AGN would allow reionisation completion to be
consistent with neutral fraction and optical depth constraints.
This means that we are effectively studying an optimal case for
reionisation by AGN, since those fainter than $M_{1450} > -18$
might already be a part of the galaxy population as discussed
in~\citet{char17}, due to the overlap between the 
galaxy and AGN luminosity functions at this faint limit. Applying a magnitude cut-off would suppress the
ionising emissivity and further delay reionisation.

\begin{figure}
\centering
\setlength{\epsfxsize}{0.5\textwidth}
\centerline{\includegraphics[scale=0.4]{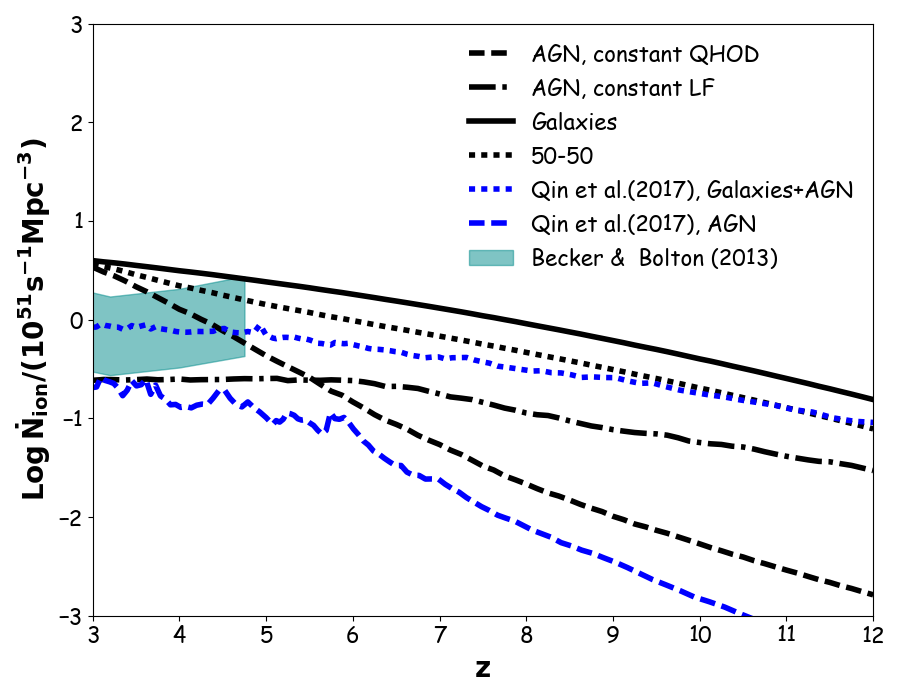}}
\caption{The comoving integrated ionising emissivity as a function of redshift. The~\citet{bec13} measurements are shown with the cyan shaded area (1-$\sigma$ level).   The emissivity evolution in our  Galaxies model (black solid) and  Galaxies+AGN model (dotted blue) by~\citet{qin17} is similar with an amplitude difference due to these models' basic framework and assumption (see text for details). The emissivity in  constant LF AGN model (dot-dashed black) grows slowly similar to galaxies driven-EoR models, which reflects the poor assumption of using a fixed AGN LF through all times. Our fiducial  constant QHOD AGN model (dashed black) produces reasonable emissivity evolution with a steep decline towards high redshift, consistent with the  AGN model (dashed blue) by~\citet{qin17}.}
\label{fig:Nion}
\end{figure} 

In summary, we have described our procedure to obtain the ionising
emissivity of AGN as a function of halo mass, and then populate AGN
into halos within {\sc SimFast21} via constraining the halo occupancy
of AGN (QHOD) using the G15 AGN LF.  The total emissivity
is calibrated to match that observed by~G15 at $z=5.75$,
which fixes our free parameter relating black hole mass to halo
circular velocity.  Our fiducial model assumes a constant QHOD,
and we will also consider a constant AGN LF.  We now study the
predictions of our model for reionisation observables, and compare
the results with our previous {\sc SimFast21} models where we
considered only star-forming galaxies.

\section{EoR Observables}\label{sec:impacts}

\subsection{{\sc SimFast21} runs}

We run all of our EoR realizations using the Time-integrated
model~\citep{hassan17} to establish a proper comparison between
the different source models. Using the same density field and halo
catalogs generated in a box size L$=300$ Mpc and N$=560^{3}$ number
of cells, we run 4 different EoR models based on different ionisation
sources as follows (and summarized in Table~\ref{models_tab}):
 \begin{itemize}

\item {\bf Galaxies}: This model only considers ionising photons
emitted by star-forming galaxies using Equation~\eqref{eq:nion} with
parameters: $\fescs$ = 0.25 , $A_{\rm ion}$ = 4.27$\times 10^{39}$,
$C_{\rm ion}$ = 0.44. These parameters are suggested by our recent
MCMC analysis in~\citet{hassan17} to match simultaneously various
EoR constraints including the SFR densities at several redshift
bins as compiled by~\citet{bouw15}, integrated ionising emissivity
at z $\sim$ 5 by~\citet{bec13} and~\citet{planck16} optical depth.

\item {\bf constant QHOD}: This is our fiducial AGN model in
which the AGN are the only source for ionising radiation using
$\fesca$=1.0, and our QHOD fitting function (Equation~\ref{eq:hod})
computed from~G15 LF fit at $z=5.75$.

\item {\bf 50-50}: This model contains an equal contribution from
the {\bf Galaxies} and {\bf constant QHOD} models, specifically
we use $\fescs$ = 0.125 and $\fesca$ = 0.5.

\item {\bf constant LF}: This is our alternative AGN model which
uses the actual LF fit of~G15 at $z=5.75$ to compute the
number of AGN at all redshifts, with $\fesca$=1.0.
\end{itemize}

\begin{table*} \LARGE
 \scalebox{0.7}{\begin{tabular}{ l  c  c  l c c}\hline
   {\bf Model }& {$ \fescs$} &  {$ \fesca$} & {\bf ionisation rate} & $\tau_e$ & $z_{\rm reion}$ \\ \hline \hline
    {\bf Galaxies} & 0.25 & 0.0 & $\fescs\, \rst $ & 0.057  &  7.5 \\ \hline 
    {\bf constant QHOD} AGN & 0.0 & 1.0 & $\fesca\, \ragn$  & 0.036 & 5.0\\ \hline 
    {\bf 50-50} & 0.125 & 0.5 & $\fescs\, \rst  + \fesca\, \ragn$ & 0.049 & 6.5\\ \hline 
    {\bf constant LF} AGN & 0.0 &  1.0 & $\fesca\, \ragn$ & 0.048 & 4.0\\ \hline
\end{tabular}}
\caption{Summary of models considered in section~\ref{sec:impacts} to compare between the AGN and star-forming galaxies impacts on different reionisation constraints. Columns (from left to right) are: models' names, the photon escape fractions from star-forming galaxies $\fescs$ and AGN $\fesca$, the ionisation rate used in Equation~\eqref{eq:new_con}, the optical depth $\tau_e$, and reionisation redshift $z_{\rm reion}$ defined at neutral fraction limit $x_{\rm HI} < 10^{-3}$.}\label{models_tab}
\end{table*}

\subsection{Ionising emissivity}

We begin by comparing the integrated ionising emissivity $\dot{N}_{\rm
ion}$ of these models, which is the total number of ionising photons
per second per comoving volume. We compare our models with results
from ~\citet{qin17} based on the DRAGONS simulation, which uses the
{\sc Meraxes} semi-analytic galaxy formation model built upon the
{\sc Tiamat} $N$-body simulation.  The \citet{qin17} model is able
to track the growth of central super-massive black holes and reproduce
wide range of observations including the observed quasar LF
from $z\sim 0.6-6$. Their model predicts that AGN contribution to
EoR is minimal, so it is interesting to compare our AGN emissivity
with theirs.

Figure~\ref{fig:Nion} shows the redshift evolution of the comoving
integrated ionising emissivity $\dot{N}_{\rm ion}$, in units of
$10^{51}$~s$^{-1}$Mpc$^{-3}$, for our four models.  For the
\citet{qin17} models (blue lines), we show their full galaxies+AGN
model as well as their AGN-only contribution.  At $z\la 5$, we show
the observational constraints from \citet{bec13} inferred from
Ly$\alpha$ forest measurements.

Our Galaxies model skirts the upper limit of \citet{bec13} constraints
while simultaneously calibrated to reproduce the \citet{bouw15} SFR
and \citet{planck16} optical depth, as discussed in~\citet{hassan17}.
Adopting a mildly redshift-dependent or even mass-dependent $\fescs$
would permit a better match with the amplitude and flat redshift
dependence of the \citet{bec13} emissivity measurements, as suggested
by~\citet{mut16}, without much altering the Thomson optical depth.

Our fiducial constant QHOD model shows a much more rapid growth
of ionising emissivity with time than the Galaxies model, which
matches the \citet{bec13} ionising emissivity at the upper end of the
observed redshift range but overshoots the low end.  In this model,
the AGN contribution overtakes the Galaxies contribution at $z\sim
3$, which is in agreement with what is typically
inferred~\citep[e.g.][]{hama12}.

We further see that our 50-50 model (dotted black in Figure~\ref{fig:Nion})
is similar and much closer to the Galaxies model than to the
constant QHOD model.  This indicates that the contribution from
star-forming galaxies dominates the ionising emissivity while AGN
contribution is minor.

Finally, the constant LF model shows a relatively shallow
evolution, approximately parallel to the galaxies case but significantly
lower in amplitude.  This falls just below the ionising emissivity
data at $z\la 5$.

The blue dotted and dashed lines show the evolution of the emissivities
from the \citet{qin17} model, with the latter showing only the
contribution from AGN.  Their AGN contribution shows a similar
redshift dependence to our constant QHOD model, which further
supports the validity of this assumption, at least down to $z\sim
6$.  It is possible that our assumption breaks down at $z\la 5$ as
their AGN contribution flattens, and if ours did this then the
agreement with the observed evolution of the emissivity would
improve.  However, DRAGONS also predicts that galaxies dominate the
ionising photon budget at all redshifts, which may be contrary to
studies of the hardness of the ionising background in the $z\sim
2-4$~IGM~\citep[e.g.][]{schaye07,Opp09}.  

The emissivity evolution in our  Galaxies model is similar to
that from the  Galaxies+AGN model by~\citet{qin17}, whereas
there is a difference in the amplitude due to these models' differences
in the $\fesc$ treatment (constant versus redshift-dependent). As
previously noted, the  constant LF AGN model shows a slowly
growing emissivity similar to those of our  Galaxies model and
 Galaxies + AGN model by~\citet{qin17}.

In summary, our fiducial  constant QHOD AGN model matches
the \citet{bec13} emissivity measurements reasonably well, at least
at $z\sim 5$, but shows a dramatically different redshift evolution
compared to our Galaxies and constant LF models.  The \citet{qin17}
DRAGONS model shows an evolution during reionisation that is
consistent with our constant QHOD model for their AGN component,
and with our Galaxies model for their overall emissivity (which is
dominated by galaxies), but somewhat lower in amplitude in each
case.  All these models are broadly in agreement with the emissivity
measures at $z\sim 5$ given current uncertainties.

\subsection{Global ionisation history and optical depth}

We now explore our model predictions for other current observational
constraints on the global evolution of reionisation, particularly
the evolution of the volume-weighted neutral fraction $x_{\rm HI}$ and
the integrated Thomson optical depth to electron scattering to the
CMB $\tau_e$.

\begin{figure}
\centering
\setlength{\epsfxsize}{0.5\textwidth}
\centerline{\includegraphics[width=3.8in, height=2.8in]{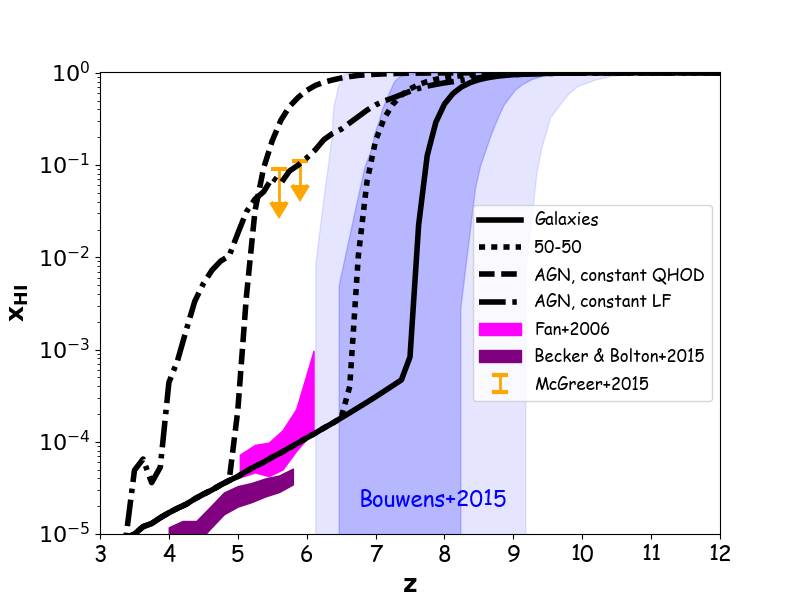}}
\caption{The volume-weighted averaged neutral fraction evolution as a function of redshift. The shaded areas: magenta, purple, and light blue are from Ly-$\alpha$ forest measurements by \citet{fan06},~\citet{bec15}, and several AGN and ly-$\alpha$ constraints (1-$\sigma$ and 2-$\sigma$) as compiled by~\citet{bouw15a}, respectively. We also compare to the model independent upper limits by~\citet{mcg15} (orange errorbars) using Ly-$\alpha$ and Ly-$\beta$ forest. It is evident that Galaxies (solid) and 50-50 (dotted) models are consistent with all observations, which implies the importance of including  star-forming galaxies to match with observations. Our AGN models,  constant LF (dot-dashed) and  constant QHOD (dashed), both complete reionisation very late. This indicates that AGN contribution to cosmic reionisation is minor.}
\label{fig:xhi}
\end{figure}

Figure~\ref{fig:xhi} shows a comparison between our models in terms
of their global ionisation history, as characterised by the
volume-weighted average neutral fraction $x_{\rm HI}$.  We see that our
 Galaxies and  50-50 models are consistent with several
Ly-$\alpha$ forest measurements (shaded areas from ~\citealt{fan06,bec15,bouw15a} and orange upper limits by~\citealt{mcg15}).
While both are consistent with data, the  50-50 model delays
reionisation by $\Delta z \sim 1.0$, owing to the fact that galaxies
are the main driver of reionisation as discussed in the previous
section and their contribution has been halved in this model.

Turning to our AGN-driven reionisation models, we see that both the
 constant LF and constant QHOD reionise the
Universe very late, with $\rm {x_{\rm HI}} < 10^{-3}$ not occurring
until $z\sim 4$ and $z\sim 5$, respectively.  This is highly
inconsistent with the~\citet{bouw15a} constraints as seen in
Figure~\ref{fig:xhi}, as well as direct observations of the Ly$\alpha$
forest in quasars at these epochs~\citep{fan06,bec15}.  The constant
LF AGN model starts reionisation earlier than constant QHOD AGN
model due to its higher emissivity at high redshifts
(Figure~\ref{fig:Nion}). This too-late reionisation strongly suggests
that AGN are unable to drive the bulk of reionisation, when
constrained to match the observed ionising emissivity after the end
of reionisation ($z\sim 5$).

As mentioned earlier, all models use the Time-integrated ionisation
condition (Equation~\ref{eq:new_con}) to identify the ionised regions
in the excursion set-formalism. We showed in~\citet{hassan17} that
this ionisation condition results in a more sudden reionisation as
opposed to our previous Instantaneous model~\citep{hassan16} which
yields an extended reionisation scenario. From Figure~\ref{fig:xhi},
all models yield a fairly sudden reionisation, consistent with our
previous results, except the  constant LF AGN model that shows
an extended reionisation.  This is because the fixed LF likely
over-estimates the number of AGN at high redshifts, and furthermore
the source population does not grow in concert with the growing sink
population, which delays reionisation.
\begin{figure}
\centering
\setlength{\epsfxsize}{0.5\textwidth}
\centerline{\includegraphics[scale=0.5]{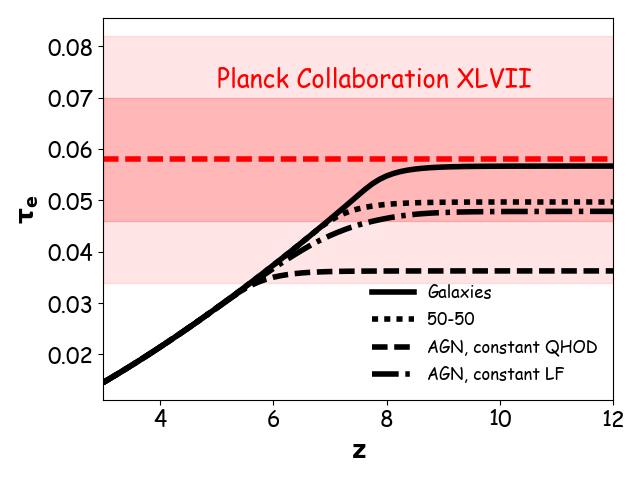}}
\caption{Thomson scattering optical depth evolution as a function of redshift. The shaded red dark and light areas represent the 1-$\sigma$ and 2-$\sigma$ levels of the recent~\citet{planck16} optical depth measurement whereas the dashed red horizontal line marks the measured {\it Planck} value ($\tau  = 0.058 \pm 0.012$). The  Galaxies model (solid) is consistent with the actual optical depth value. The  50-50 and constant LF models obtain a lower optical depth of $\tau \sim$ 0.049, matching the lower 1-$\sigma$ level of planck. Our fiducial  constant QHOD AGN models produces a very low optical ($\tau \sim$ 0.036) which lies at the lower limit of 2-$\sigma$ level.}
\label{fig:tau}
\end{figure}  

Figure~\ref{fig:tau} shows the evolution of the Thomson scattering
optical depth ($\tau_e$) as a function of redshift in these models.
The Galaxies model is consistent with the recent~\citet{planck16}
measurements (red shaded areas), mostly because it was constrained to do so
via MCMC. The  50-50 and  constant LF yield a
lower optical depth of about $\tau \sim$ 0.049, consistent with the
lower limit of 1-$\sigma$ level. However, the optical depth obtained
by our fiducial  constant QHOD AGN model is very low ($\tau
\sim 0.036$) at the lower limit of the 2-$\sigma$ level.
Essentially, this model does not produce enough early photons in
order to obtain a sufficient ionised path length to the CMB.

One might question why \citet{maha15} was able to match these
constraints based on the G15 model and thereby argue for purely
quasar-driven reionisation, whereas we reach an opposite conclusion.
There are two main differences.  First, \citet{maha15} made a rough
fit to the ionising emissivity, which resulted in a much flatter
redshift dependence than we obtain from our constant QHOD model,
more like our constant LF model except higher in amplitude by $\sim\times 2-3$
(see Figure~\ref{fig:epsilon912}).
Indeed, our constant LF model results in an ionising emissivity
evolution much like theirs, and is consistent (albeit marginally)
with $\tau_e$.  Second, \citet{maha15} purely relied on counting
emitted photons, and did not account for an evolving
spatially-clustered nature of sinks which absorbs more photons, and
hence retards reionisation particularly at early epochs compared
to a spatially-homogeneous clumping factor model. These differences
result in substantially more neutral gas at early epochs in our
model, and lead to strong disagreement with the measured $\tau_e$
as well as $x_{\rm HI}$.

Our 50-50 model can plausibly match the constraints within current
$1\sigma$ uncertainties.  Recall that for our Galaxies model,
\citet{hassan17} found $\fesc=0.25^{+0.26}_{-0.13}$, which means
that the galaxy contribution alone in the 50-50 model corresponds
to an escape fraction that is at the $1\sigma$ bound allowed by the
MCMC fit.  Correspondingly, the predicted $\tau_e$ and the redshift
of reionisation are near the $1\sigma$-low end of their respective
allowed observational ranges.
\begin{figure*}
\centering
\setlength{\epsfxsize}{0.5\textwidth}
\centerline{\includegraphics[width=8in, height=6in]{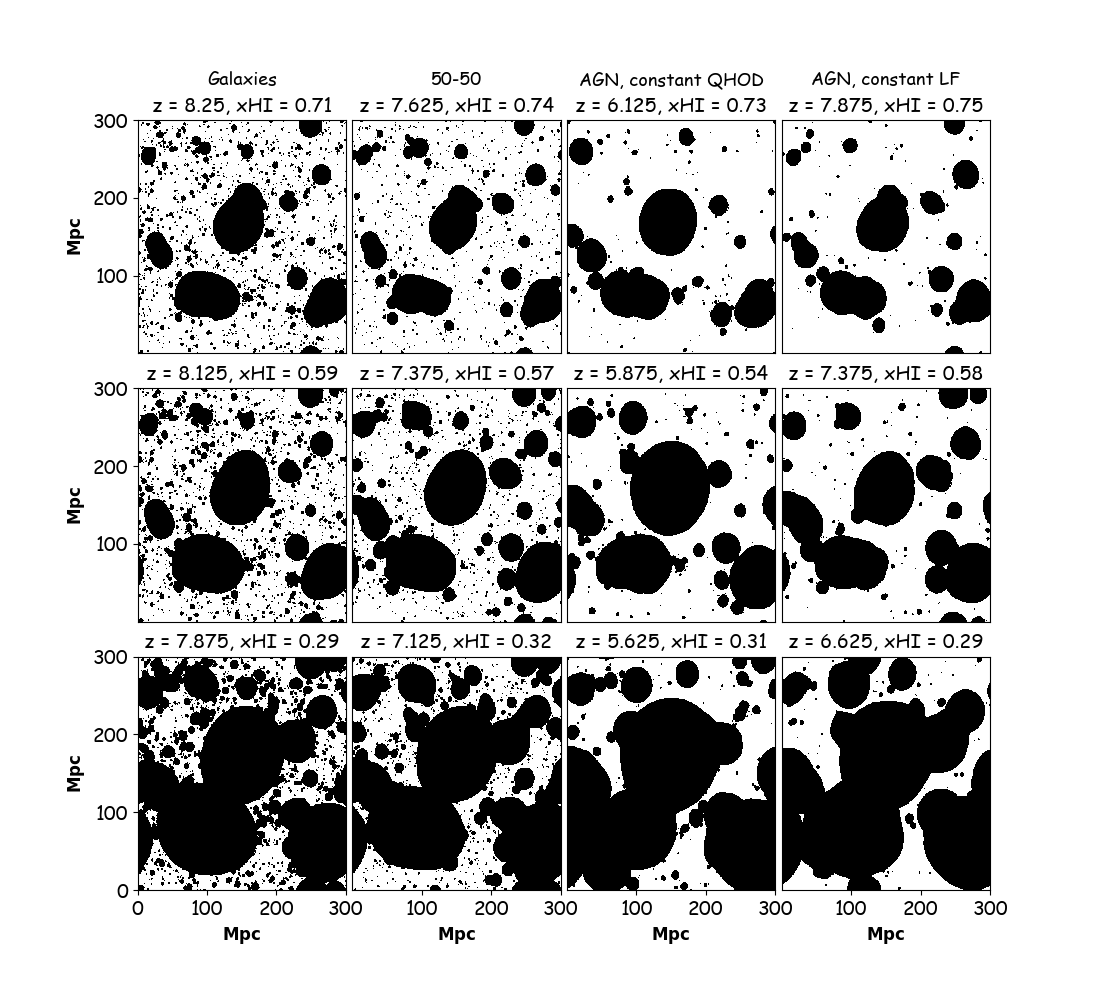}}
\caption{Slices of the ionisation box each of a size $300\times300\times0.535$~Mpc$^{3}$ from our four models at different stages of reionisation. Top to bottom: x$_{HI} \approx 0.25, 0.5 , 0.75$. Left to right:  Galaxies,  50-50, constant QHOD,  constant LF models. White and black represent neutral and ionised regions respectively. The  constant QHOD and constant LF AGN models show large HII bubbles as compared to  Galaxies model. The  Galaxies and  50-50 models show similar topology, indicating that galaxies play stronger role in determining the HII regions properties. Actual redshifts and neutral fractions are quoted on top of each map.}
\label{fig:maps}
\end{figure*} 
The \citet{qin17} AGN model yields an optical depth of $\tau \sim
0.025$, corresponding to an end of reionisation at $z \sim 3$. Our
 constant QHOD AGN model would obtain similar results if a
magnitude cut-off had been implemented in the ionising emissivity
integral to exclude the very faint AGN.  Even with our more favorable
case of reionisation by AGN, reionisation still occurs very late.

Overall, we find that our constant QHOD model that is constrained
to match the AGN emissivity of G15 can adequately match the global
ionising emissivity at $z\sim 5$, but strongly fails to reionise
the Universe by $z\sim 6$ and produces too low Thomson optical
depth.  A constant LF model does somewhat better at matching
constraints, but the underlying assumption is not physically
well-motivated, does not match the observed emissivity evolution
from $z\sim 6\rightarrow 4$, and is inconsistent with the self-consistent
calculations of \citet{qin17}.  A 50-50 model is still within the
allowed bounds of the observations considered here, but any larger
contribution from AGN would be disfavoured -- and we reiterate that our
AGN model is already pushed towards increasing the AGN emissivity
as much as possible.  Our $x_{\rm HI}$ and $\tau_e$ constraints for
all these models are listed in Table~1.  Our results thus suggest
that AGN-dominated reionisation is highly unlikely, and therefore
that galaxies dominate the ionising photon budget during the EoR.

\subsection{EoR topology}\label{subsec:eortop}

We have shown that our AGN-only models are highly disfavoured given
current observational data.  However, a 50-50 model is still
permissible, if only marginally.  Clearly, increasing the precision
of current measures should in principle enable more stringent
constraints on the relative contribution of AGN and star-forming
galaxies to reionisation.  But we can also appeal to other aspects
such as the topology of reionisation in order to discriminate between
models. This will be particularly fruitful in the era of 21cm EoR
experiments, which will quantify the power spectrum of neutral
hydrogen on large-scales.  In this section, we discuss the topology
of neutral gas in our various models, and in the following section
we will quantify this by forecasting the 21cm power spectrum for
upcoming experiments.

We first investigate these models' differences in terms of their
ionisation field maps. We choose to compare these models' maps at
a fixed neutral fraction, since we have shown in~\citet{hassan17}
that the 21cm fluctuations are more sensitive to the topology of
the ionisation field while the density field contribution is
secondary. This then allows us to compare the topology even though
the actual redshift where a given ionisation occurs varies substantially
between models.

Figure~\ref{fig:maps} shows 2D maps of the ionisation field of all
models at different neutral fractions (${x_{\rm HI}} \approx 0.25,
0.5, 0.75$), projected through the entire volume.  Note that the
redshifts at which this occurs are much later for the AGN-only
models, consistent with their late reionisation.
\begin{figure*}
\centering
\setlength{\epsfxsize}{0.5\textwidth}
\centerline{\includegraphics[scale=0.5]{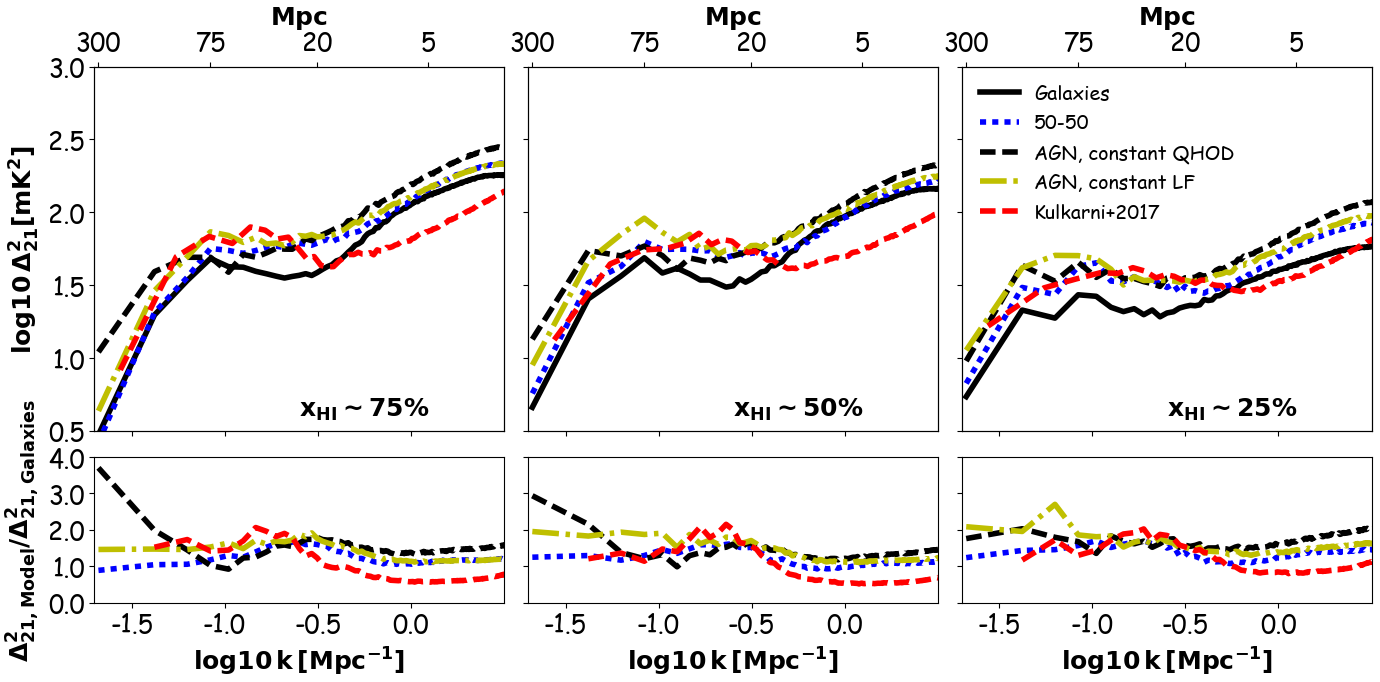}}
\caption{Top: the 21cm power spectrum comparison between all models at fixed neutral fractions.  We also compare to the recent AGN model developed by~\citet{kul17} (dashed red). The  Galaxies (black solid) and  50-50  (dotted blue) models produce similar power spectra on small and large scales, particularly at intermediate  (${x_{\rm HI} \sim 0.5}$) and early (${ x_{\rm HI} \sim 0.75}$) stages of reionisation. Likewise, the constant QHOD (dashed black) and  constant LF (dot-dashed yellow) AGN models yield similar power spectra. The AGN models produce power spectra that is higher by a factor of $\sim \times$ 1.5$-$2 than the  Galaxies model. The power spectra obtained by~\citet{kul17} AGN model agree with our AGN models relatively on large scales but they are lower by a factor of $\sim \times$ 2$-$2.5 on small scales, due to these models' differences in populating AGN into halos (see text for details). Bottom: the ratio of each model power spectrum to that of Galaxies as function of the scale k, which shows the difference between each model to Galaxies.}
\label{fig:pk21}
\end{figure*} 
Our Galaxies model shows a range of bubble sizes, with many small
bubbles around low-mass galaxies that have low clustering.  The
largest bubbles towards the end of reionisation span $\sim 100$~Mpc,
in agreement with many previous studies~\citep[e.g.][]{bar04,furl04,mesinger11,zah11,ian14,maj14}.  It is clear that there
will be significant power on all scales owing to this topology.

The  constant QHOD and constant LF AGN models show fairly
similar \ion{H}{ii} bubble sizes and distributions.  Compared to the
Galaxies case, there are fewer small bubbles owing to the fact that
AGN tend to populate more massive halos than the typical galaxy
contributing to reionisation in the Galaxies model.  However, there
are still some small halos hosting small bubbles even in the AGN
case.  This contrasts with the~\citet{kul17} AGN model where AGN
are assigned only in the massive halos using a circular velocity
cut-off, and thus their model does not yield any small-scale \ion{H}{ii} regions
(see their Figure~2).  Nonetheless, because the duty cycle of AGN
in our model is very low in low-mass halos, many fewer small bubbles
are seen compared to the Galaxies case.

For large ionisations, the ionisation maps of the  constant QHOD and
constant LF AGN models display larger \ion{H}{ii} regions as compared
with those of  Galaxies model.  This is necessary to compensate for
the lack of numerous small bubbles, in order to achieve a similar
neutral fraction.  This is driven by the strong AGN clustering as
suggested by the input G15 LF at its faint end.

The 50-50 model is, not surprisingly, intermediate between the
Galaxies and AGN-only models.  The small bubbles are less prominent
owing to the lower $\fescs$.  The maximum bubble sizes are comparable
to but slightly larger than in the Galaxies model.  Hence the
star-forming galaxies tend to drive the topology even when substantial
AGN are present.  This comes from the facts that each halo mass bin
(magnitude bin) includes fewer AGN than the number of possible
hosting halos as implied by the AGN duty cycle estimates (e.g.
\citealt{shan09}), and hence star-forming galaxies would overcome
the impact of those fewer AGN on the ionisation field topology.
The difference between AGN (constant QHOD and constant LF) and
star-forming galaxies (Galaxies and perhaps 50-50) dominated models
increases as reionisation proceeds and becomes clear at late stage
of reionisation (bottom row for ${x_{\rm HI} \sim 25\%}$ of
Figure~\ref{fig:maps}). We expect these trends to be reflected
quantitatively in their 21cm power spectra.
\begin{table*} \LARGE
 \scalebox{0.65}{\begin{tabular}{ l  l  c  c c}\hline
   {\bf Experiment } & Design &  Diameter [m] & Collecting area [m$^2$] & Receiver temperature [mK] \\ \hline \hline
    {\bf LOFAR} & 48 tiles of bow-tie high band antennae  & 30.75  & 35,762 & 140,000    \\ \hline 
    {\bf HERA} & 331 hexagonally packed antennae & 14 & 50,953  & 100,000  \\ \hline 
    {\bf SKA} & 866 compact core antennae & 35 & 833,189 & 100 T$_{\rm sky}$ + 40,000  \\ \hline 
\end{tabular}}
\caption{Summary of parameters used in {\sc 21cmSense} package to obtain the thermal noise sensitivity for each experiment. Columns (from left to right) are: experiments' names, designs, antenna diameter, total collecting area, and the reciver temperature. The sky temperature is given by T$_{\rm sky}$ = 60$\lambda^{2.55}$ K.}\label{array_tab}
\end{table*}
\subsection{The 21cm power spectrum}

Using the ionisation fields of these models at fixed neutral fractions
(Figure~\ref{fig:maps}), we now compute our key EoR observable,
namely the 21cm power spectrum. Assuming that the spin temperature
is much higher than the CMB temperature, the 21cm brightness
temperature takes the following form:
  \begin{equation}\label{21cm}
\delta T_{b} (\nu) = 23 x_{\rm HI}\Delta \left( \frac{\Omega_{b} h^{2}}{0.02} \right) \sqrt{  \frac{1+z}{10} \frac{0.15}{\Omega_{m}h^{2}}} \left( \frac{H}{H + d v/dr} \right) \mathrm{mK},
\end{equation}
where $d v/dr$ is the comoving gradient of the line of sight component
of the peculiar velocity. Using this equation, it is straightforward
to create the 21cm brightness temperature boxes from which we compute
the 21cm power spectrum as follows: $\mathrm{\Delta^{2}_{21} \equiv
k^{3}/(2\pi^{2}\, V) < | \delta T_{b} (k)  |^{2}_{k} >}$.

In top panels of Figure~\ref{fig:pk21}, we show a comparison between
our models' predicted 21cm power spectrum, as well as with the
AGN-dominated models of~\citet{kul17}.  Bottom panels show the ratio
of each model 21cm power spectrum to the Galaxies model, in order
to clearly display the models differences as a function of the scale
$k$.  Consistent with the ionisation maps (Figure~\ref{fig:maps}),
the  Galaxies model has less power than the AGN-only models, which
at a fixed $x_{\rm HI}$ are themselves rather similar.  This shows that
the 21cm power spectrum is somewhat insensitive to the method by
which we populate AGN (constant QHOD versus constant LF),
even though there is a clear difference in the reionisation histories
due to differences in the ionising emissivity evolutions (see
Figures~\ref{fig:Nion}~\& ~\ref{fig:xhi}).  Relative to the Galaxies
case, at early epochs the differences with the AGN-only models peaks
at intermediate scales (k $\sim$ 0.3 $-$ 0.1), typical of the AGN
bubble sizes.  At later epochs, however, there is not much scale
dependence to the variation, and the AGN models are simply about a
factor of $1.5-2\times$ higher than the Galaxies model, owing to
the stronger clustering of G15 AGN observations.

The 50-50 models yield similar 21cm power spectra both on small
and large scales to the Galaxies case at the early ($x_{\rm HI}
\sim 0.75$) and intermediate (${x_{\rm HI} \sim 0.5}$) stages of
reionisation. This confirms our previous finding that star-forming
galaxies are dominant in determining the ionised regions properties
and hence their associated 21cm fluctuations during the early stages
of reionisation.  At later stages, there is an increasing difference
between the 50-50 model and the Galaxies case, owing to the increased
contribution of AGN at later epochs.  Hence it may be optimal to
look at the latter stages of reionisation in order to obtain more
quantitative constraints on the fractional contribution of AGN.

\begin{figure*}
\centering
\setlength{\epsfxsize}{0.5\textwidth}
\centerline{\includegraphics[scale=0.5]{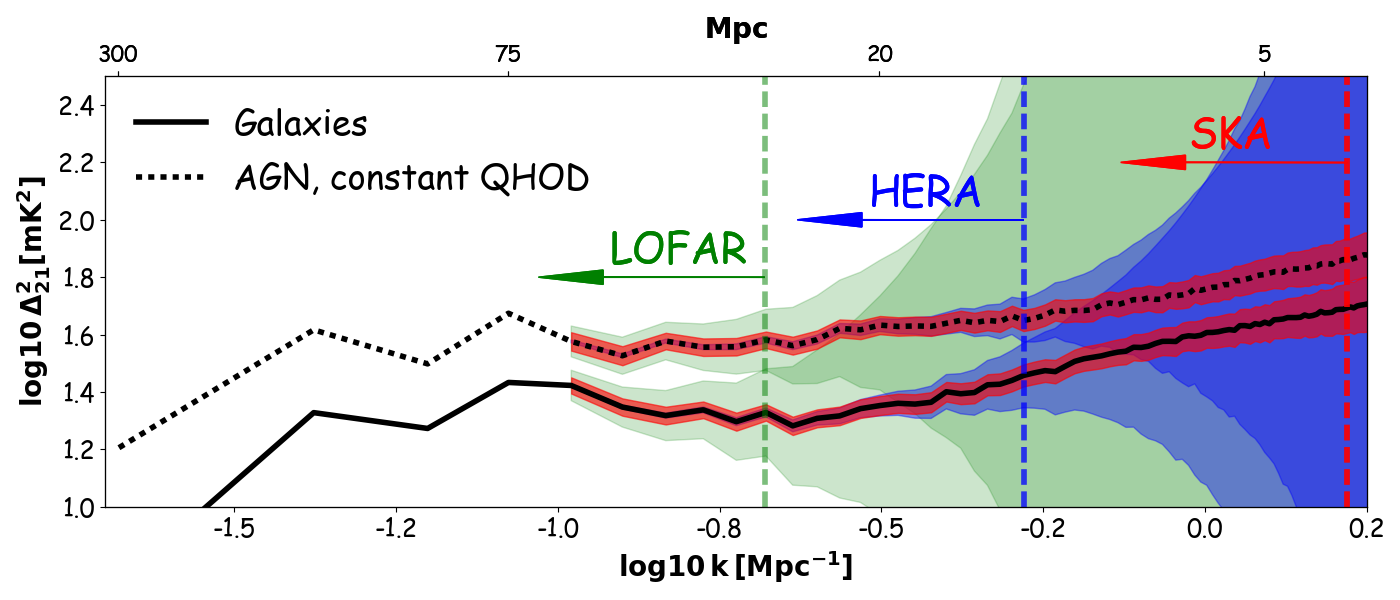}}
\caption{Mock 21cm EoR power spectrum observations using the same telescope designs and configurations at z = 8  and ${x_{\rm HI}} \sim$ 30$\%$ from our  Galaxies (black solid) and  constant QHOD  AGN (black dashed, tuned to match  Galaxies model optical depth $\tau$) models. Shaded areas show the 1-$\sigma$ erorrbars obtained using {\sc 21cmSense} package for our constructed EoR arrays: SKA (red), HERA (blue), LOFAR (green). Vertical dashed lines represent the scale at which a specific experiment may distinguish between these models (the scale at which errorbars overlap from a specific experiment). Future 21cm observations by these experiments will be able to discriminate between these models at their corresponding sensitivity limits.}
\label{fig:21cmerrors}
\end{figure*} 
\begin{figure*}
\centering
\setlength{\epsfxsize}{0.5\textwidth}
\centerline{\includegraphics[scale=0.5]{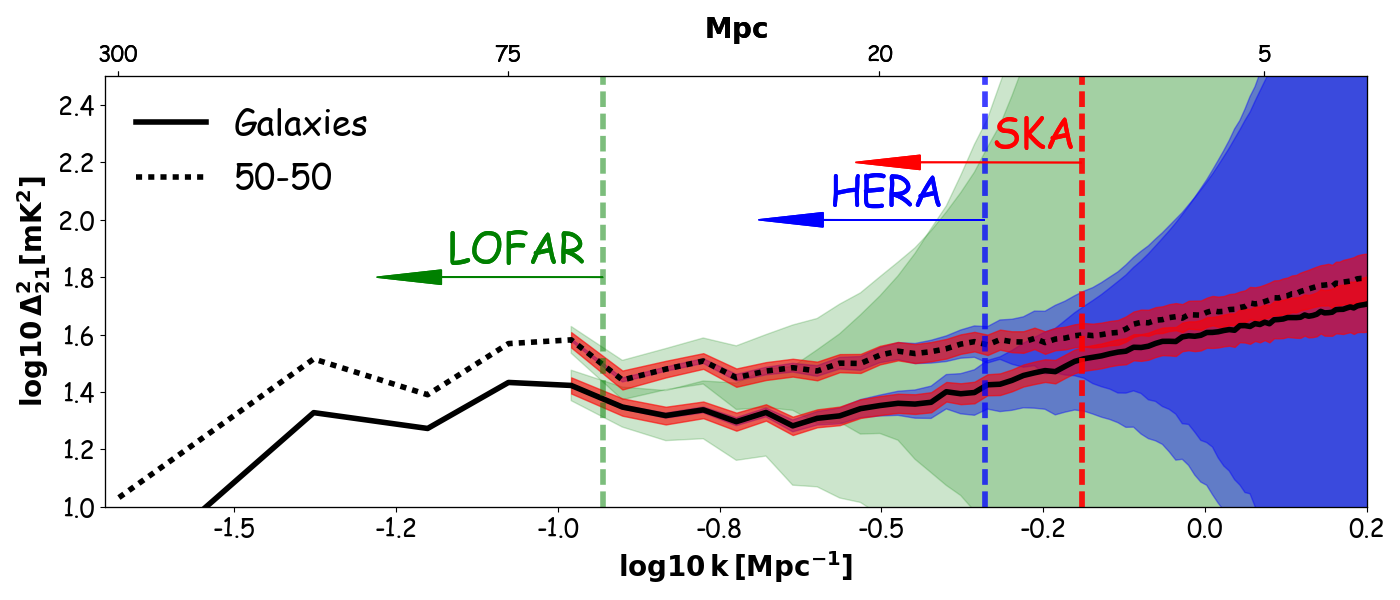}}
\caption{Mock 21cm EoR power spectrum observations using the same telescope designs and configurations at z = 8  and ${x_{\rm HI}} \sim$ 30$\%$ from our  Galaxies (black solid) and  50-50 (black dotted, tuned to match  Galaxies model optical depth $\tau$) models, similar to Figure~\ref{fig:21cmerrors}. Only future 21cm observations by HERA and SKA will be able to discriminate between these models at their corresponding sensitivity limits.}
\label{fig:21cmerrors_50}
\end{figure*} 

Our AGN models agree with the~\citet{kul17} AGN model reasonably
well for the large-scale 21cm power spectrum, while their small
scale power is suppressed by a factor of $\sim \times 2-2.5$ relative
to ours. As mentioned before in~\S\ref{subsec:eortop}, our AGN model
yields small-scale ionised regions owing to populating AGN randomly
at all halo mass bins (magnitude bins) into their hosting halos
using the actual number of AGN as suggested by the~G15 LF observations,
as opposed to a unity duty cycle in massive halos in the~\citet{kul17}
AGN model.  By constraining our LF to that observed, our AGN model
predicts a non-unity duty cycle, which occasionally populates small
halos with AGN and thus boosts the 21cm power spectrum on small
scales.

\subsection{Forecasting 21cm power spectra to constrain AGN models}

Our work has shown that AGN driven-EoR models are photon-starved,
in agreement with many others, and as such plausible AGN models are
unlikely to fully drive reionisation.  Nonetheless, a substantial
AGN contribution such as in our 50-50 model is still allowable given
current data, which begs the question, will future 21cm data provide
more stringent constraints on the contribution of AGN to reionisation?

To answer this question, we focus our analysis on three different
low-frequency radio interferometer designed to measure the 21cm EoR
power spectrum: the Low Frequency Array
(LOFAR)\footnote{http://www.lofar.org/}, the Hydrogen Epoch of
Reionisation Array (HERA)\footnote{http://reionization.org}, and
the Square Kilometer Array
(SKA-Low)\footnote{https://www.skatelescope.org}. 

To forecast the power spectra for these facilities, we use the same
recipe presented in~\citet{hassan17}, outlined as follows: We first
select the redshift (observed frequency) at which we compute the
21cm power spectrum for each model. To establish a proper comparison, we operate these three array designs, with parameters summarized in Table~\ref{array_tab}, in a drift-scanning mode for 6 observing hours per day for 180 days at 8 MHz bandwidth.  We then add the total uncertainty
which includes the thermal noise and sample variance using the {\sc
21cmSense}\footnote{https://github.com/jpober/21cmSense}, a package for calculating the sensitivity of 21cm
experiments to the EoR power
spectrum.  We refer
to~\citet{par12} for basics of the radio interferometer sensitivities,
to~\citet{pob13,pob14} for more details on observation strategies
and foreground removal models, and to~\citet{hassan17} for the
experiments designs and configurations. We only change
the foreground removal method to use the optimistic model developed in~\citet{pob14}
in which the foreground wedge extends to the Full-Width Half-Max (FWHM) of the experiments'
primary beam. This will extend our analysis to cover more large scales than using a
moderate model in which foreground wedge extends only to 0.1 $h$ Mpc$^{-1}$
beyond horizon limit. 

From Figure~\ref{fig:pk21}, we notice that the large variations
between the  Galaxies and constant QHOD AGN models occurs at
the late stages of reionisation (${x_{\rm HI} \sim 25 \%} $) when
the HII bubbles begin to overlap. We thus create our 21cm mock
observations at these epochs. Since the reionisation occurs very
late in the constant QHOD AGN model (see Figure~\ref{fig:xhi}),
we re-tune the model (using $A=1.25\times 10^{7}$, which we note
would substantially overproduce the ionising emissivity constraints)
to match the optical depth as obtained by  Galaxies model. We then
perform our 21cm mock observations at $z=8$ and ${x_{\rm HI}}\sim
0.3$ for both models in order to conduct a comparison at the same
redshift and neutral fraction.

Figure~\ref{fig:21cmerrors} shows a 21cm mock observation comparison
at z=8 (${x_{\rm HI}}\sim 0.3$) between  Galaxies and the re-tuned
 constant QHOD AGN models.
The shaded area shows the total uncertainties (1-$\sigma$ level) expected from the
experimental designs summarised in Table~\ref{array_tab} using the {\sc
21cmSense} package. The vertical lines mark the scale at which the
1-$\sigma$ errorbars from a specific experiment overlap, corresponding to the
sensitivity limit for each experiment where the models can be
distinguished.

Given that it can only probe relatively large scales, LOFAR will
have some difficulty discriminating between the Galaxies and
constant QHOD models, as they lie within $\sim$ 2-$\sigma$ of each
other. However, HERA should be able to distinguish between these
models rather well on scales above about 10~Mpc, and the larger
baselines of SKA-low will enable discrimination to significantly
smaller scales.  From Figure~\ref{fig:21cmerrors}, we see that
LOFAR, HERA, and SKA can discriminate between these models during
the latter stages of reionisation at scales of $k<0.21$ Mpc$^{-1}$
($>30$ Mpc), $k<0.53$ Mpc$^{-1}$ ($>12$ Mpc) and $k<1.66$ Mpc$^{-1}$
($>4$ Mpc), respectively.

The fact that HERA and SKA can easily discriminate AGN-only and
Galaxies-only models suggests that it may be possible to constrain
the fractional contribution of AGN using such data. 
We thus repeat the same steps above, except now for the 50-50 model,
tuned to match Galaxies $\tau_e$.

Figure~\ref{fig:21cmerrors_50} shows a forecasting comparison between
the Galaxies and 50-50 models.  Since the 50-50 model yields
21cm power amplitude that is closer to Galaxies than constant
QHOD model does, the scales at which experiments overlap are shifted
towards large scales (compare vertical dashed lines in
figure~\ref{fig:21cmerrors} versus \ref{fig:21cmerrors_50}). This
shows that LOFAR is unlikely to discriminate between these models,
unless a very optimistic foreground removal is applied to detect
the signal on large scales ($k<0.12$ Mpc$^{-1}$ ($>53$ Mpc)), which
are highly contaminated by foregrounds~\citep{pob14}.  Given a
successful foreground removal, HERA and SKA can discriminate between
the Galaxies and 50-50 models during the latter stages of reionisation
at scales of $k<0.46$ Mpc$^{-1}$ ($>14$ Mpc) and $k<0.65$ Mpc$^{-1}$
($>10$ Mpc), respectively as shown by vertical dashed lines in
Figure~\ref{fig:21cmerrors_50}.

Note that we have shown 1-$\sigma$ uncertainties, which
is unlikely to be sufficient to robustly discriminate between
Galaxies and AGN models.  If one instead requires 3-$\sigma$ to
distinguish between the models, then LOFAR fails to distinguish
between our models, but HERA and SKA are still successful albeit
on scales somewhat larger than those obtained with the 1-$\sigma$
limit.  

Figure~\ref{fig:21cmerrors} \& \ref{fig:21cmerrors_50} both illustrate
how future 21cm observations could potentially help constrain the
nature of the source population.  While current observations cannot
rule out the 50-50 model, in principle HERA and SKA should be able
to do so straightforwardly, assuming they can reach their target
sensitivities.  Until these facilities come online, ancillary
observations will continue to improve.  Hence comprehensive models
that are able to make predictions for, and constrain to, a wide
variety of EoR data are vital for optimizing the scientific information
extracted from future observations including 21cm data.

\section{Conclusion}\label{sec:con}

We have presented predictions for the 21cm power spectrum arising
from AGN-driven reionisation models, and contrasted them
with predictions from galaxy-driven models and models with a mixture
of sources. The AGN source population is placed into galaxy halos
using a physically motivated prescription based on the~G15
AGN observations, deriving a Quasar Halo Occupancy Distribution of
AGN at $z=5.75$ from this and using it to evolve the number of
AGN to higher redshifts.  This framework is implemented into our
Time-integrated version of {\sc SimFast21}, which self-consistently
accounts for recombinations and the evolution of structure.  We
have calibrated these AGN models to reproduce ionising emissivity
constraints, and compared them with models in which ionising radiation
is dominated by star-forming galaxies. Our key findings are
as follows:

\begin{itemize}
\item When tuned to match the~G15 ionising emissivity
constraints, AGN-only models produce very late reionisation at $z\ll
6$ (Figure~\ref{fig:xhi}).  If we assume a constant halo occupancy
for AGN as is consistent with other observational constraints and
models, then the predicted Thomson optical depth is only 0.036,
well below Planck constraints (see Figure~\ref{fig:tau}).  This strongly disfavours
AGN as providing the dominant source for reionising photons.

\item We determine a quasar halo occupancy distribution (QHOD) that
is near unity for very massive halos, but drops to sub-percent level
for more typical halos.  This is directly interpretable as a duty
cycle for AGN.  This also explains why reionisation is so late in
this AGN-only model, because AGN populate massive halos more
frequently and hence their emissivity contribution grows strongly
only at late epochs (Figure~\ref{fig:Nion}), thereby not ionising
enough volume to match the optical depth measurements.

\item Our results are consistent with those from the AGN-only models
by~\citet{qin17} using the DRAGONS semi-analytic models, who also
found that AGN could not dominate reionisation.  Our Galaxies model
is also in broad agreement with their full model, supporting their
result that star-forming galaxies can provide sufficient photons
as a function of redshift to reionise.

\item A model where we assume a constant AGN luminosity function
at $z\geq 5.75$ can barely match the Planck $\tau_e$, but still
reionises very late, and moreover it is not physically well motivated
and disagrees with the measured evolution of the AGN luminosity
function at $z\sim 4-6$~\citep{gia15}.  It does, however, result in a
global emissivity evolution similar to that assumed in \citet{maha15},
but even in this case we do not confirm their result that such a
model is viable, likely because we include the effects of recombinations
along with a more sophisticated accounting of the clustering of
sources and sinks.

\item Our AGN-only model produces larger HII bubbles as compared
with our Galaxies-only model (see Figure~\ref{fig:maps}), consistent
with results from another semi-numerical model by~\citet{kul17}
(Figure~\ref{fig:maps}).  This results in a larger 21cm power
spectrum amplitude by $\sim\, \times\, 1.5-2$ as compared with that
from the galaxies-only model (see Figure~\ref{fig:pk21}).

\item We examine a model which includes a 50\% contribution
from galaxies and AGN (assuming a constant QHOD).  We find that
this model can barely satisfy current $\tau_e$ and $x_{\rm HI}$
constraints.  At early epochs, the galaxies contribution dominates the 
power spectrum, but during the latter stages of reionisation the
quasars contribution is more significant, and the power spectrum
deviates more substantially from the galaxies-only case.

\item Future 21cm observations by LOFAR, HERA and SKA can discriminate
between the constant QHOD AGN and Galaxies models
during late reionisation at scales of $k<0.21$ Mpc$^{-1}$ ($>30$
Mpc), $k<0.53$ Mpc$^{-1}$ ($>12$ Mpc) and $k<1.66$ Mpc$^{-1}$ ($>4$
Mpc), respectively (see Figure~\ref{fig:21cmerrors}).  HERA and SKA
will also be able to distinguish between our 50-50 and galaxies-only
models, and thus potentially constrain the fractional contribution
of AGN to reionisation.

\item We have assumed an optimistic model for the AGN photon output
rate by extrapolating to very low luminosities ($M_{1450}
> -18$).  There are also suggestions that G15 over-estimates the
number of high-$z$ AGN and thus their total
emissivity~\citep[e.g.][]{parsa17}.  In either scenario, our claim
that AGN cannot dominate reionisation is further strengthened.
\end{itemize}

It might still be possible that our AGN-only models could
match all EoR observations simultaneously, if we relax some of these
models' assumptions. For instance, our AGN-only models
depend on two parameters, namely, the photon escape fraction $\fesca$ and 
the black hole formation efficiency $A$ (see Equation~\ref{eq:mbh}). 
These results are already obtained using 100\% $\fesca$, and it is clear
that those models would reionise much earlier if we adopt $\fesca \gg 100\%$,
which is not physical. The $A$ parameter has been tuned to allow these models
to reproduce the G15 data at z=5.75. If we choose not to calibrate these models 
to match the G15 data, we then can adopt larger $A$ which results in a smaller 
QHOD and earlier reionization. For example, the QHOD in the most faint magnitude
bin at z=5.75 decreases from $\sim 2.7\%$ to $1.7\%$ as the $A$ increases from 
$5\times 10^{5}$ to $10^{6}$. This shows that allowing the AGN-only models to
form more efficient black holes (large $A$) result in a fewer number (less QHOD)
of them since the QHOD quantifies the fraction of active halos with AGN. From
Figure~\ref{fig:xhi}, we see that the reionization in the constant LF AGN model (with $A=10^{6}$)
starts earlier than in the constant QHOD AGN model (with $A=5\times 10^{5}$), but because there is no
evolution in the source population (fixed LF), the model reionises much later than the constant QHOD.
It is then clear that if we adopt larger $A$ to form more effeicent black holes, our AGN-only models will
yield an early reionization. In this case, these models might produce consistent
reionization histories and optical depths as compared with the observations. Given
the large uncertainties in the ionising emissivity measurements by G15 and~\cite{bec13},
these models might still be consistent with these measurements' 2-$\sigma$ level at $A > 10^{6}$.
We expect that the AGN clustering remains unaffected at a specific neutral fraction for larger $A$ values
and hence the future 21cm observations can still discriminate between our AGN-only versus Galaxy-only models.
The AGN-only models also assume that $M_{\rm bh} - v_{\rm cir}$ correlation (Equation~\ref{eq:mbh}) is valid even 
at high redshifts. This adds more uncertainties since such a correlation has only been measured in the local universe.
More physically motivated and self-consistent AGN modelling at high redshift is clearly required to understand their formation and evolution as a function of redshift. Analogously to our stellar source model, one may allow the black hole emissivity to scale super-linearly with black hole mass ($\ragn \propto M^{C}_{\rm bh}$). This new parameter $C$ would indeed affect the AGN clustering and the corresponding ionised bubble sizes, as previously seen in our star-forming galaxies models ($\rst \propto M^{C}_{\rm h}$). We leave investigating this dependence for future works.

Improved high redshift AGN observations are clearly desirable in
order to more robustly determine the AGN luminosity function,
particularly at the low-luminosity end.  The results of such
observations can be incorporated straightforwardly into the framework
we have presented here, and will provide better constraints on the
contribution of AGN to reionisation.  Our framework illustrates
that the QHOD is an effective approach to evolve the number of AGN during
the EoR.  Our analysis indicates that despite there being potentially
more faint AGN than previously believed, star-forming galaxies still
dominate the neutral gas topology and ionising photon budget.

\citet{dalo16},~\citet{mit16}, and~\citet{ono17} have
all independently demonstrated that AGN-only models overheat the
IGM, inconsistent with Ly-$\alpha$ temperature measurements~\citep{bec11},
due to the early onset of He II reionization.~\citet{fin16} further
have shown that these models also over-ionise the metals when
compared with observed metal absorption line measurements~\citep{dod13}.
Our findings corroborate these results via a different approach.

AGN could still be important for reionisation because of their
long-range heating effects owing to their harder emission spectrum,
as well as for setting the shape of the metagalactic ionising flux
that is important for interpreting metal-line absorption
data~\citep[e.g.][]{fin16}.  Early AGN are in and of themselves
interesting in order to understand the emergence of supermassive
black holes particularly at early epochs.  Our results here suggest
that future 21cm experiments will have a key role to play in
constraining the amount of AGN activity and its contribution to the
metagalactic flux during the EoR.

 \section*{acknowledgements}
The authors acknowledge helpful discussions with Jonathan Pober and
Jonathan Chardin. We thank Girish Kulkarni and Yuxiang Qin for
making their models' data available to us.  SH is supported by the
Deutscher Akademischer Austauschdienst (DAAD) Foundation.  RD and
SH are supported by the South African Research Chairs Initiative
and the South African National Research Foundation. MGS is supported
by the South African Square Kilometre Array Project and National
Research Foundation.  Part of this work was conducted at the Aspen
Center for Physics, which is supported by National Science Foundation
grant PHY-1066293.  RD acknowledges long-term visitor support
provided by the Simons Foundation's Center for Computational
Astrophysics, as well as the Distinguished Visitor Program at Space
Telescope Science Institute, where some of this work was conducted.
Computations were performed at the cluster ``Baltasar-Sete-Sois'',
supported by the DyBHo-256667 ERC Starting Grant, and the University
of the Western Cape's ``Pumbaa" cluster.

\bsp	
\label{lastpage}
\end{document}